# Intermolecular coupling and fluxional behavior of hydrogen in phase IV

Alexander F. Goncharov[1,2,*], Irina Chuvashova[2], Cheng Ji[3], Ho-kwang Mao[3]

[1]Key Laboratory of Materials Physics and Center for Energy Matter in Extreme Environments, Institute of Solid State Physics, Chinese Academy of Sciences, Hefei, Anhui 230031, China

[2]Geophysical Laboratory, Carnegie Institution of Washington, 5251 Broad Branch Road, Washington, District of Columbia 20015, USA

[3]Center for High Pressure Science and Technology Advanced Research, Shanghai 201203, China

*corresponding author e-mail: agoncharov@carnegiescience.edu

**We performed Raman and infrared (IR) spectroscopy measurements of hydrogen at 295 K up to 280 GPa at an IR synchrotron facility of SSRF. To reach the highest pressure, hydrogen was loaded into toroidal diamond anvils with 40 μm central culet. The intermolecular coupling has been determined by concomitant measurements of the IR and Raman vibron modes. In phase IV, we find that the intermolecular coupling is much stronger in the graphene (G) like layer of elongated molecules compared to the $Br_2$ like layer of shortened molecules and it increases with pressure much faster in the G layer compared to the $Br_2$ layer. These heterogeneous lattice dynamical properties are unique features of highly fluxional hydrogen phase IV.**

Dense hydrogen demonstrates a number of fascinating phenomena [1], and theory predicts even more spectacular behaviors at higher pressures, which remained to be explored [2, 3]. Of particular interest is a behavior related to an increase of the kinetic energy and thus quantum atomic motion, which may lead to a change in the character of the chemical bonds [4] or even to a decline in the melting temperature, which can ultimately result in a liquid ground state [5, 6]. Hydrogen with the lightest atoms manifests the most suitable system to explore such effects. However, reaching the appropriate states requires very high pressures (~1 TPa), which remains technically challenging.

Static high-pressure techniques have been recently progressing aggressively stimulated by scientific goals of better understanding materials under extremes (*e.g.* in planetary interiors), competition with dynamic compression techniques, and new advances in first principles calculations. High-pressure molecular hydrogen $H_2$ is expected to transform to a metallic monatomic state at high pressures [7, 8], but the route remained unclear. Until 2012, only three molecular phases of $H_2$ have been widely recognized: plastic (fully orientationally disordered) hcp phase I and orientationally ordered phases II at low temperature and III at low temperature and high pressure. While phase II, possessing quantum ordering features [9, 10], is unusual for molecular crystals, phase III was thought to be similar to common orientationally ordered phases in other classical molecular crystals such as diatomics [11]. Assuming that this behavior continues to higher pressures, one could expect that higher-pressure $H_2$ polymorphs would be classically ordered molecular crystals as the DFT theory predicts [11]. However, experiments revealed an unusual "mixed molecular and atomic" phase (predicted theoretically in Ref. 11) to crystallize at 225 GPa on compression of phase I via phase III at 295 K [4]. This phase (originally determined as *Pbcn*) was found to be slightly less stable than *Cmca-4* in DFT calculations [11-13], however recent diffusion quantum Monte Carlo calculations showed that *Pc*-48 phase IV is more stable at close to room temperatures [14]. The structure of phase IV (Fig. 1) consists of alternating layers of two types: graphene-like (G), where elongated molecules form a honeycomb-like lattice, and another quasi- hexagonal layer, where the molecules are shortened ($Br_2$) and also oriented close to the basal plane.

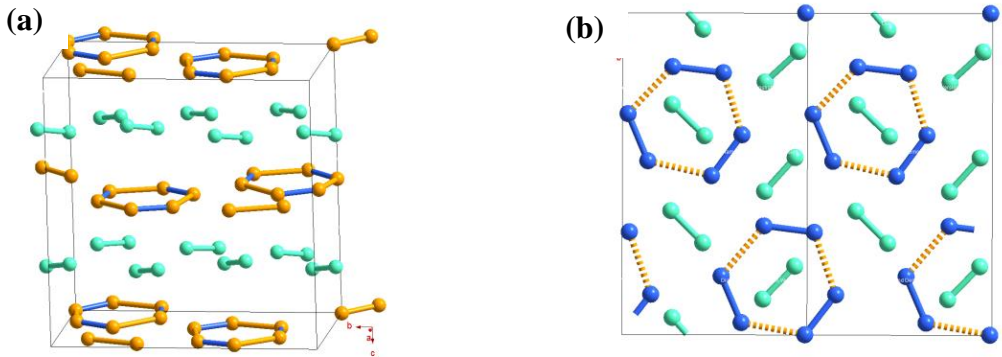

Fig. 1. (a) The crystal structure of phase IV of hydrogen (*Pc*-48) after theoretical predictions [13]. The G-layers contain the elongated molecules (blue intramolecular bonds) associated in quasi-hexagons (yellow intermolecular bond), while the $Br_2$ layer consists of shortened molecules (green bond). In the panel (b) only two layers are shown projected along the *c*-axis.

However, this classical structural picture has been questioned in molecular dynamics simulations [15-18], showing that the structure of this new phase is highly dynamic, where the atoms in the elongated molecules of the G layer show a diffusive motion, which can be viewed as the rotation of a three-molecule ring and an even longer atomic migration. Thus, the chemical bonds change their space location migrating with time yet preserving the local lattice symmetry at each time; we call this behavior fluxional in analogy with that of some molecular substances at ambient conditions [19]. On the other hand, the shortened molecules in the $Br_2$ layer are orientationally disordered but the atoms do not migrate between the molecular sites. Previous optical spectroscopy investigations [4, 20-22] demonstrated that phase IV has distinct Raman and IR spectra characterized by the presence of two vibron modes (in layers G and $Br_2$); the lower frequency vibron modes ($v_1$) show a softening and broadening behavior under pressure which is consistent with the molecules in the G layer to be very short lived [16]. In contrast, the higher frequency vibron ($v_2$) corresponds to shortened molecules in the $Br_2$ layer. However, Raman and IR experiments have been performed separately making it difficult to compare the results, especially concerning the splitting between the IR and Raman modes. Here, we report the results of concomitant IR/Raman experiments on the same sample yielding the reliable information about the intermolecular coupling in both G and Br$_2$ layers. We find that the intermolecular coupling in the two types of layers are very different, with that in the G layer being much stronger and increasing more rapidly with compression.

We performed the experiments at a newly constructed system at the beamline BL01B of the Shanghai Synchrotron Radiation Facility [23, 24]. The system combines synchrotron Fourier-Transform (FT-IR) spectroscopy with a broadband laser visible/near infrared (IR) and conventional laser Raman spectroscopy in one instrument. The all-mirror custom confocal IR microscope was used in a transmission mode and the FT-IR spectra were recorded with a mercury-cadmium-telluride (MCT) detector with the 0.05 x 0.05mm$^2$ crystal dimensions in the spectral range of 700-10000 cm$^{-1}$. The synchrotron beam diameter was about 10 μm measured in the near IR spectral range [24]. The transmission IR measurements were performed in a single-channel mode using the IR spectrum of the same sample measured at the condition where IR absorption was relatively weak (*e.g.* in phase I) [25] as the reference. The same IR optics was used to measure visible and near IR transmission spectra with the Raman spectrometer equipped with array CCD (200-1100 nm) and InGaAs (900-1650 nm) detectors (Princeton Instruments PIXIS and NIRvana, respectively). The Raman microscope objective lens (Mitutoyo 50 X, NA=0.4) is interchangeable with the IR reflective objectives diverting the optical path to the Raman spectrometer. A 660 nm single-line solid-state laser was used to excite the Raman spectra in a back scattering geometry and the signal was analyzed using three narrow bandpass holographic notch filters and a single grating 500 mm focal length spectrograph equipped with a CCD detector. Raman and IR experiments were performed at 295 K at the same nominal pressure measured with Raman spectra of the stressed diamond [26] and finely corrected using the spectral position of the main Raman vibron band as presented in Ref. 27.

Two (out of a dozen attempts) diamond anvil cell experiments were successful in reaching pressures in access of 200 GPa which are needed to reach phases III and IV of hydrogen at 295 K. One was performed to 240 GPa using conventional single beveled diamonds with 40 μm tip diameter. Another used toroidal anvils [28, 29] (Fig. 2) machined with FIB from conventional beveled anvils with 50 μm tip diameter, yielding a 40 μm diameter tips. These anvils were also coated (using sputtering) with alumina of 50-100 nm thickness. Rhenium was used as the gasket. Hydrogen was loaded at room temperature using a compressor pumped it up to 150 MPa. The sample dimensions were approximately 10 μm (at 160 GPa) reducing down to 6 μm at the highest pressure of 280 GPa.

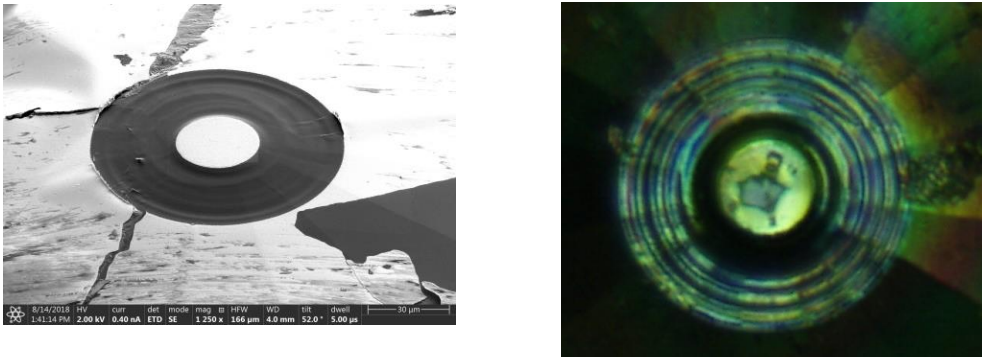

Fig. 2. Loading of hydrogen in the toroidal diamond anvils. Left panel: electronic beam image of the diamond tip machined with FIB. Right panel: optical image of the loaded hydrogen sample in the transmitted and reflected light at approximately 160 GPa.

The IR spectra of the vibron modes (Fig. 3(a)) show an increase in intensity at the transition to phase III, where one strong vibron mode is observed [30]. Above 224 GPa, this mode abruptly splits giving rise to a doublet (Fig. 4(a)). The low-frequency band softens and broadens with pressure, while the high-frequency one is almost pressure independent and is gaining the intensity. Concomitant Raman measurements at the same conditions (Fig. 3(b)) show a similar behavior of the Raman vibron band albeit the frequencies are shifted to lower energies. This energy distinction is because Raman and IR vibron modes have different vibration patterns for the molecules that belong to the same unit cell: in-phase for Raman and out-of-phase – for IR active modes. Thus, the Raman-IR splitting value represents the strength of the intermolecular coupling [25, 31]. Although the unit cell of $Pc$ phase IV consists of totally 24 H$_2$ molecules suggesting the same number of intramolecular vibron modes, only four of them are observed in Raman and IR as others have much smaller intensities [13]. The intermolecular coupling is commonly represented as Van Kranendonk's hopping matrix elements $\varepsilon_{ij}$ [32], that correspond to intermolecular coupling strengths. The difference in the Raman and IR vibron frequency in hcp phase I of hydrogen is $6\varepsilon$, where $\varepsilon/2$ corresponds to the pair interactions between nearest neighbor molecules. Under pressure, the intermolecular coupling normally increases representing a normal tendency for compression of molecular crystals, where intermolecular distances contract much faster than intramolecular (these can even expand) due to heterogeneity of the interatomic interactions. This behavior has been established for phases I, II and III of hydrogen, but phase IV reveals an anomalous behavior as elaborated below.

Our combined concomitant IR-Raman experiments allow determining the splitting of the vibrational band independently of the pressure measurements, which could complicate the previous attempts (cf. Ref. 22), where such determination has been performed based on separate IR and Raman measurement, in which pressure could be determined with a large uncertainty (*e.g.* Ref. 27) (Fig. 4(a)). Here, based on our improved measurements, we come to a different conclusion than in Ref. 22: we find a strong increase with pressure in the vibrational coupling in the G-layer of phase IV. Furthermore, we deduced the intermolecular coupling in the G- and Br$_2$ layers using a simple nearest-neighbor model, where the molecules in the Br$_2$ layer have six nearest neighbors in the same layer (cf. twelve in phase I), while there are only four nearest molecule in the G-layer (Fig. 1 and Fig. S2 in Ref. 33). For simplicity, we assumed all equal intermolecular couplings of the same kind, which can be tentatively supported by the dynamical nature of phase IV, where one can expect time averaging of the bond lengths. The interlayer couplings between molecules of different kinds do not contribute into the mode splitting, as the corresponding vibron modes are decoupled [12,13]. Within this simple model, the results (Fig. 4(b)) show a nearly continuity in the intermolecular coupling for $Br_2$-layers with quasi-hexagonal layers of phase III and its strong increase for G-layers through the III-IV phase transition. At these conditions (≥270 GPa) the whole inter- intra-molecular bonding concept is about to break down because the difference between intra- and inter-molecular bond strength becomes much less substantial. Moreover, the molecules in the G layer are short living as they decompose and recombine within a picosecond time scale; this dramatically increases anharmonic effects [16]. On the other hand, the intermolecular coupling in $Br_2$ layers is much weaker demonstrating an intriguing bonding distinction between G and $Br_2$ layers, which also results in a charge transfer and band gap opening [16] stabilizing the structure. Our experiment demonstrate that the bandgap is still open in phase IV up to at least 280 GPa (Fig. S1 in Ref. 33) in agreement with previous observations [4]. Concerning the intermolecular bonding anisotropy, it can be explained naïvely as due to a difference in the intramolecular bond lengths in G and $Br_2$ layers which leaves a complementary length for the intermolecular bonds in these adjacent layers. Thus, phase IV of hydrogen manifests a Peierls distortion of some kind. Vibrational spectroscopy represents a unique way of probing this unusual high-pressure behavior capturing a local atomic configuration to which it is very sensitive.

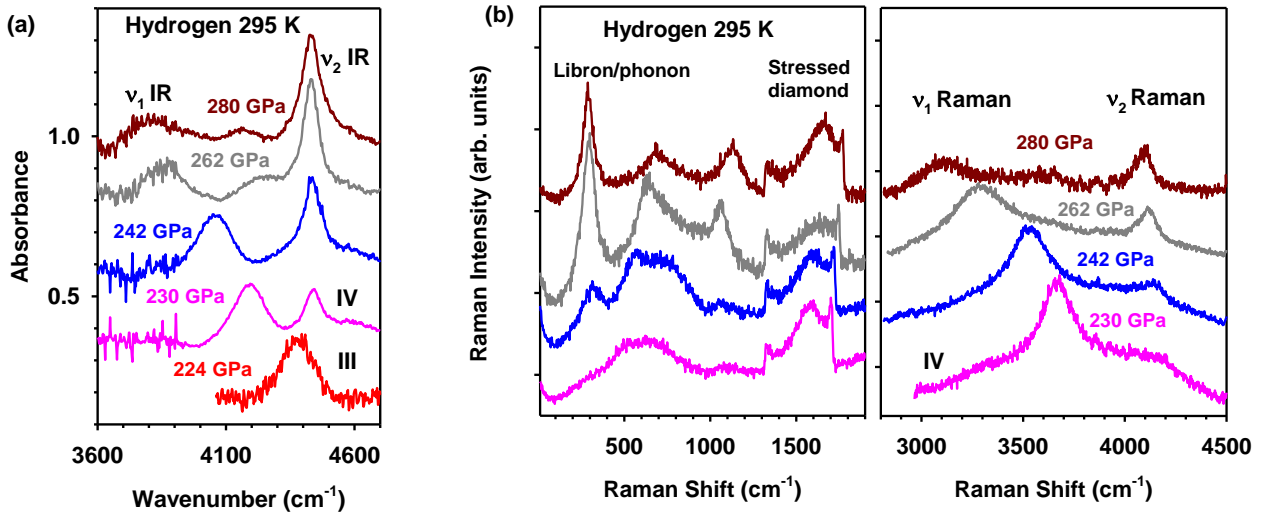

Fig. 3. Synchrotron IR and Raman spectroscopy in compressed hydrogen up to 280 GPa at 295 K. (a) IR spectra of the vibron modes in phases III and IV. (b) Raman spectra of phase IV as pressure increases; left and right panels show the libron and phonon modes and the vibron modes, respectively. The vibron modes denoted as $\nu_1$ and $\nu_2$ correspond to the strongly and weakly bounded G and $Br_2$ layers, respectively. The results are in a qualitative agreement with previous measurements [4, 6, 20-22].

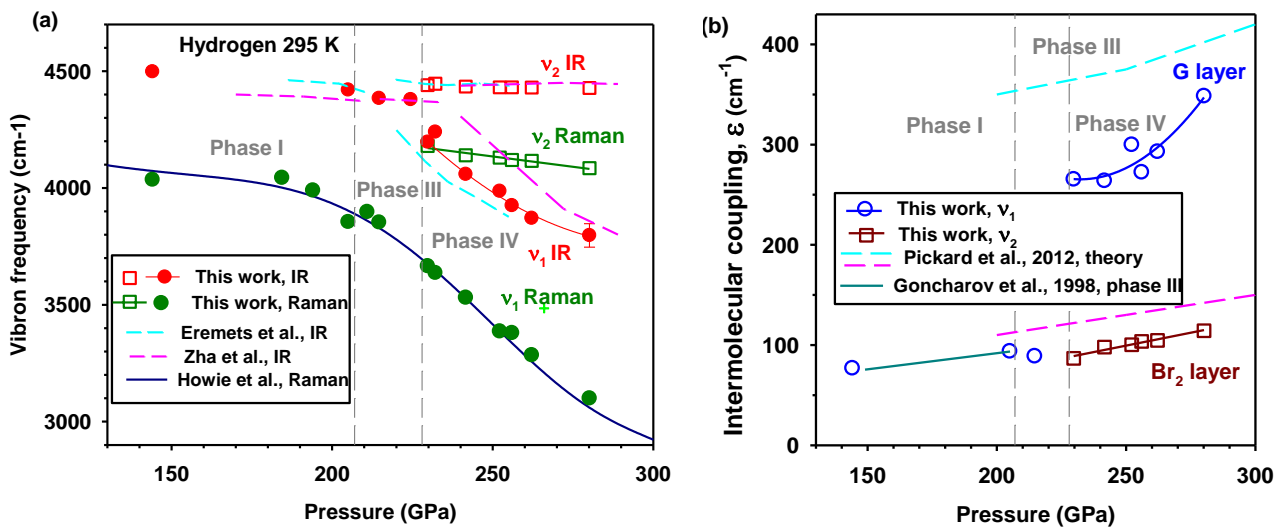

Fig. 4. (a) Raman and IR vibron frequencies as a function of pressure: symbols - this work; pressure is determined via Raman measurements of the stressed anvils and further corrected to match the position of the main Raman vibron mode $\nu_1$ according the calibration of Ref. 27; an uncertainty of the frequency determination is smaller than the size of the symbols except for the last pressure point, where an asymmetric $\nu_1$ IR peak was observed; the IR results of Ref. 21 agree fairly with Zha et al. [22] and are not shown; (b) intermolecular coupling calculated from the difference in the IR and Raman vibron frequencies (see text for details): symbols- from this work and dashed lines- theoretically calculated [13]; the results below 205 GPa are compared to previously measured in phase III [34].

The unique properties of phase IV of hydrogen demonstrated here allow us to speculate on how this phase can be probed by other techniques (*e.g.*, X-ray diffraction, XRD). Unlike vibrational spectroscopy, XRD captures an averaged over the time positions of the atoms (via scattering on the electrons). Determination of a detailed structure, that includes knowledge of the bond lengths, for fluxional crystals such as phase IV, would require careful single-crystal structure analysis, which is beyond the current technical capabilities. However, using the theoretical structural predictions, we can model the expected XRD of phase IV and compare this with the results of the experiments that are currently feasible [35]. To describe a highly diffusive G layer, we assumed that it has a graphene structure, which corresponds to an extreme member of the family of mixed structures with *Ibam* structure (*e.g.* Ref. 36). In the $Br_2$ layer, due to its rotationally disordered state, the shortened $H_2$ molecules were approximated by the molecular centers of mass. The modeled XRD pattern (Fig. S3 in Ref. 33 is almost indistinguishable from that of an hcp $H_2$ (phase I), where the molecules again are approximated by the centers of their mass. These considerations emphasize the power of vibrational spectroscopy in determinations of structure of light fluxional molecules and suggest exercising a caution in interpreting scarce XRD data, which are currently available for hydrogen at high pressures.

We thank S. Vitale and Z. Geballe for FIB machining of the diamond anvils and Lingping Kong for help with the IR synchrotron beamline. This work was supported by the NSF EAR-1763287, the Army Research Office, the Deep Carbon Observatory, the Carnegie Institution of Washington, the National Natural Science Foundation of China (11504382, 21473211, 11674330, and 51727806), the Chinese Academy of Science (YZ201524), and a Science Challenge Project TZ201601. A.F.G. was partially supported by the Chinese Academy of Sciences Visiting Professorship for Senior International Scientists (2011T2J20) and the Recruitment Program of Foreign Experts.

# Intermolecular coupling and fluxional behavior of hydrogen in phase IV

Alexander F. Goncharov[1,2,*], Irina Chuvashova[2], Cheng Ji[3], Ho-kwang Mao[3]

[1]Key Laboratory of Materials Physics and Center for Energy Matter in Extreme Environments, Institute of Solid State Physics, Chinese Academy of Sciences, Hefei, Anhui 230031, China

[2]Geophysical Laboratory, Carnegie Institution of Washington, 5251 Broad Branch Road, Washington, District of Columbia 20015, USA

[3]Center for High Pressure Science and Technology Advanced Research, Shanghai 201203, China

This file contains supplementary Figures S1-3.

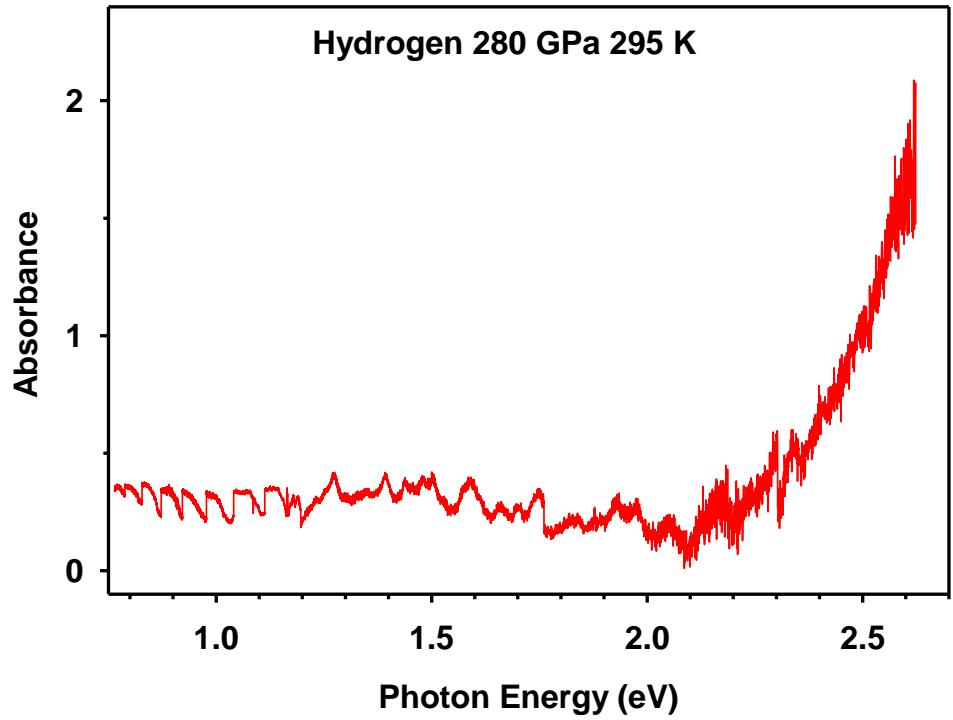

Fig. S1. Optical absorption spectrum of hydrogen at 280 GPa obtained concomitantly with IR and Raman spectroscopy measurements of Fig. 3.

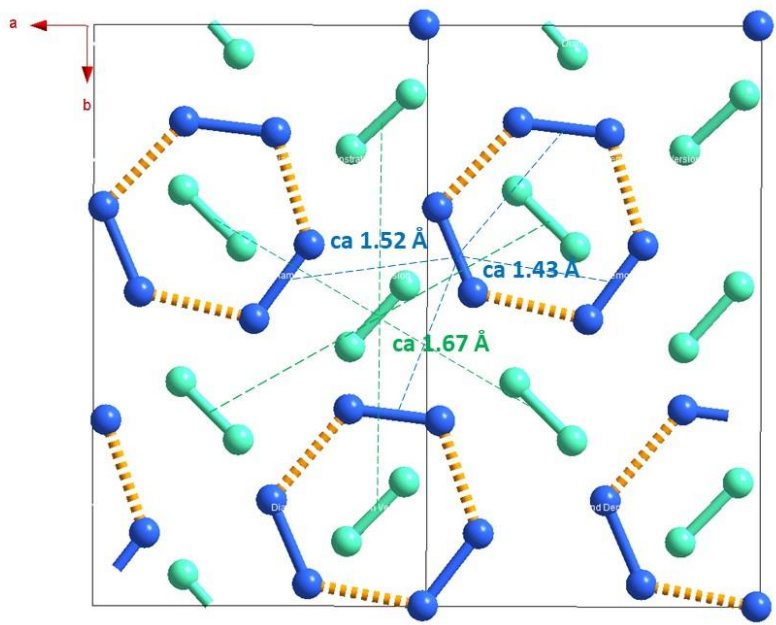

Fig. S2. The molecular arrangement in Pc hydrogen phase IV structure. As in the Fig. 1(b), only two atomic layers are shown projected along the *c*-axis. The molecules in the $Br_2$-layer (green) have six nearest almost equidistant neighbors. The molecules in the G-layer (blue) have four nearest neighbors, two closest of which belong to the same strongly intermolecular linked group of three molecules and two farthest- to the next one. Due to a large atomic motion in this layer, these distances are expected to be time averaged in the real structure.

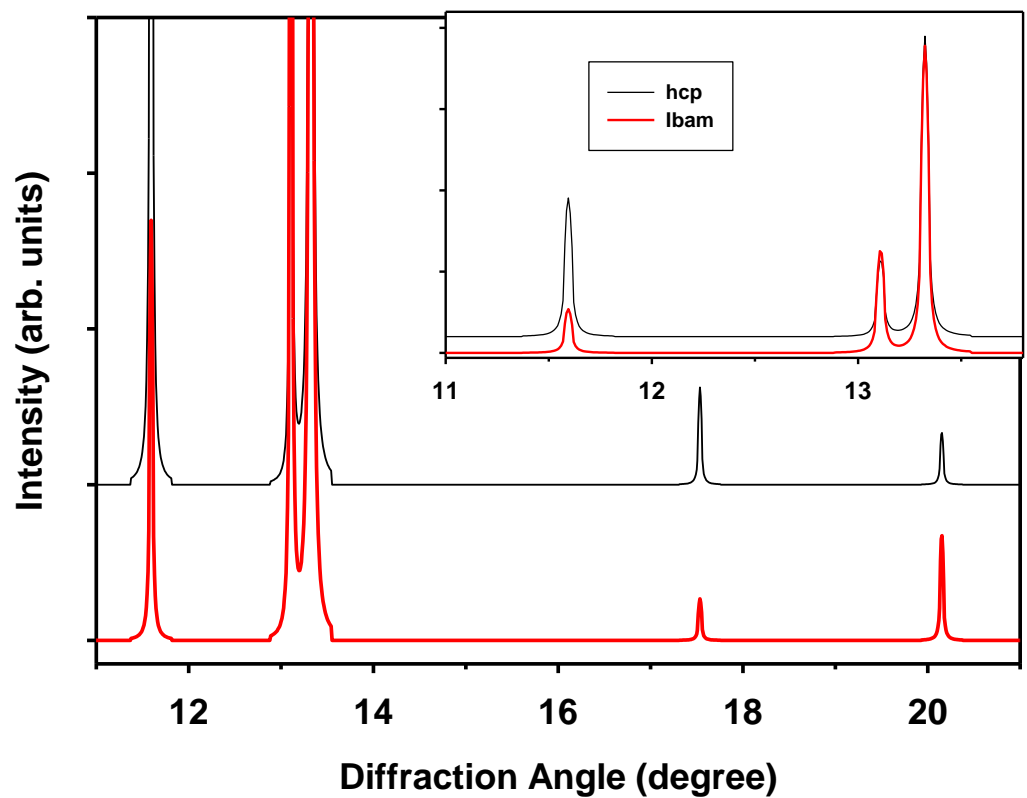

Fig. S3. Modeled XRD of hcp (phase I) and *Ibam* (prototype of phase IV) phase of hydrogen at 254 GPa. The X-ray wavelength is 0.2952 Å.